\documentclass[11pt]{article}

\usepackage{fullpage}
\usepackage{url}

\usepackage{amsmath}

\usepackage[font=footnotesize]{subfig}

\usepackage[noadjust]{cite}
\usepackage{algorithmic}
\usepackage{epsfig,graphics,graphicx,color}
\usepackage[usenames,dvipsnames]{xcolor}

\usepackage{listings}

\renewcommand{\algorithmicrequire}{\textbf{Input:}}
\renewcommand{\algorithmicensure}{\textbf{Output:}}
\renewcommand{\algorithmicrepeat}{\textbf{synchronize}}
\renewcommand{\algorithmiccomment}[1]{// #1}

\begin{document}

\title{Practical Tera-scale Walsh-Hadamard Transform}
\author{\textsc{Yi LU}\\
National Research Center of Fundamental Software, Beijing, P.R.China\\
Department of Informatics, University of Bergen, Bergen, Norway\\
Email: {dr.yi.lu@ieee.org}
}

\date{}

\lstdefinestyle{custom}{
 captionpos=t,
 belowcaptionskip=1\baselineskip,
 breaklines=true,
 frame=L,
 xleftmargin=\parindent,
 showstringspaces=false,
 basicstyle=\footnotesize\ttfamily,
 keywordstyle=\bfseries\color{green!40!black},
 commentstyle=\itshape\color{purple!40!black},
 identifierstyle=\color{black},
 stringstyle=\color{orange}
}

\lstset{style=custom}

\maketitle

\begin{abstract}
In the mid-second decade of new millennium, the development of IT has reached unprecedented new heights. As one derivative of Moore's law, the operating system evolves from the initial 16 bits, 32 bits, to the ultimate 64 bits. Most modern computing platforms are in transition to the 64-bit versions. For upcoming decades, IT industry will inevitably favor software and systems, which can efficiently utilize the new 64-bit hardware resources. In particular, with the advent of massive data outputs regularly, memory-efficient software and systems would be leading the future.

In this paper, we aim at studying practical Walsh-Hadamard Transform (WHT). WHT is popular in a variety of applications in image and video coding, speech processing, data compression, digital logic design, communications, just to name a few. The power and simplicity of WHT has stimulated research efforts and interests in (noisy) sparse WHT within interdisciplinary areas including (but is not limited to) signal processing, cryptography. Loosely speaking, sparse WHT refers to the case that the number of nonzero Walsh coefficients is much smaller than the dimension; 
the noisy version of sparse WHT refers to the case that the number of large Walsh coefficients
is much smaller than the dimension while there exists a large number of small nonzero Walsh coefficients.
Clearly, general Walsh-Hadamard Transform is a first solution to the noisy sparse WHT, 
 which can obtain all Walsh coefficients larger than a given threshold and the index positions.
 In this work, we study efficient implementations of very large dimensional general WHT. Our work is believed to shed light on noisy sparse WHT, which remains to be a big open challenge. Meanwhile, the main idea behind will help to study parallel data-intensive computing, which has a broad range of applications.

\end{abstract}

\section{Introduction}

Nowadays, we can perform the typical in-memory computing workloads
for the problem size as large as $O(2^{30})$ within minutes. 
Suppose that such workloads has the complexity linear in the size.
From now on, the performance of these workloads will be dominantly limited by
memory latency, which is on the order of 100 $ns$ currently and
grows at a slow rate of about 5.5\% per year (i.e., 
doubles every ten years).
This makes that the typical in-memory processing power tends 
to converge to an interesting critical threshold $O(2^{32})$.

Given the exponential growth of the size of demanding workloads 
and the convergence of the threshold for in-memory processing power, 
we aim at making
 practical memory-constrained system and algorithms to achieve near-optimal performance.
In this paper, we study efficient tera-scale Walsh-Hadamard Transform (WHT) \emph{for the first time}.
We intend to make it a long-term international not-for-profit project.
WHT has gained increasing research popularity and found a variety of scientific and engineering applications over the past 
(cf. \cite{Book_walsh_app,Book_walsh_app_image}).
In particular,
the topic of (noisy) sparse WHT has attracted most academia attention from various areas: 
signal processing \cite{Vetterli2015,isit2015wht,arxiv2015wht}, cryptography (and coding theory) \cite{LPN_original_paper,eprint2016serge,vaudenay_textbook2006,my_arxiv2015}. 
Informally speaking, sparse WHT refers to the case that the number of nonzero Walsh coefficients
is much smaller than the dimension;
 the noisy version of sparse WHT refers to the case that the number of large Walsh coefficients
is much smaller than the dimension while there exists a large number of small nonzero Walsh coefficients.
In signal processing, more research efforts are taken in order to 
efficiently obtain those large Walsh coefficients and the index positions
 using less number of time-domain components of the signal.
In cryptography (and coding theory),
two long-term research themes exist:
1) researchers study super-sparse WHT of super-large dimension in the noisy setting with emphasis on better time and memory complexities than general Walsh-Hadamard Transform (e.g., \cite{LPN_original_paper,eprint2016serge}).
2) researchers aim at studying practical implementation of noisy WHT with both largest possible dimension and strongest possible noise,
and somehow can tolerate with solving part of those large Walsh coefficients and index positions 
(e.g., \cite{vaudenay_textbook2006,my_arxiv2015}).
Clearly, general Walsh-Hadamard Transform is a first solution to the noisy sparse WHT, 
 which can obtain all Walsh coefficients larger than a given threshold and the index positions.
In this work, we are motivated to study
\emph{practical} Walsh-Hadamard Transform with focus on a very large signal dimension $2^n$.
Our results show that general Walsh-Hadamard Transform of dimension $2^{40}$ can be done
on the PC with 2.2GHz CPU
and 16GB RAM
 within 3 weeks using 8TB disk space.
With dimension $2^{35}$, WHT can be done with time 7 hours, 5.3 hours over the rotation disk and 
the flash disk respectively.
This compares favorably with the MPI implementation results 
\cite[Table~5, Page 109]{LPN_MPI2014} of noisy sparse WHT in cryptography (in Theme 1) when the parameter $\eta$ is large.

The rest of the paper is organized as follows.
In Section \ref{sec:1},
we give briefs on WHT; moreover, we give a formal definition on the signal-to-noise ratio (SNR),
which is very important for the interdisciplinary topic of (noisy) sparse WHT.
In Section \ref{sec:2}, we present architecture-unaware parallel WHT.
We give our results on practical large-scale external WHT for the first time in Section \ref{sec:3}; 
we consider both rotation-based disks and flash disks.
In Section \ref{sec:4}, we give more results on practical tera-scale WHT.
We give summary in Section \ref{sec:end}.

\section{Briefs on Walsh-Hadamard Transform}
\label{sec:1}

Given a real-valued function 
$f: GF(2)^n \to \rm{R}$, which is defined on an $n$-tuple binary vector of input,
the Walsh-Hadamard Transform of $f$, denoted by $\widehat{f}$, is another real-valued function defined as
\begin{equation}\label{E_def1_WHT}
\widehat{f}(i)=\sum_{j\in GF(2)^n}(-1)^{\langle i,j\rangle} f(j),
\end{equation}
for all $i \in GF(2)^n$, where $\langle i,j\rangle$ denotes the inner product between two $n$-tuple binary vectors $i,j$.
For later convenience, we give an alternative definition below.
Given an input array ${\bf x}=(x_0,x_1,\ldots,x_{2^{n}-1})$ of $2^n$ reals in the time-domain, 
the Walsh-Hadamard Transform ${\bf y} = \widehat{{\bf x}} =(y_0,y_1,\ldots,y_{2^{n}-1})$ of ${\bf x}$ is defined by
\begin{equation}\label{E_def2_WHT}
y_i = \widehat{x}_i = \sum_{j\in GF(2)^n} (-1)^{\langle i,j\rangle} x_j,
\end{equation}
for any $n$-tuple binary vector $i$. 
We call $x_i$ (resp. $y_i$) the time-domain component (resp. transform-domain coefficient or simply Walsh coefficient) 
 of the signal with dimension $2^n$.
We refer the reader to \cite{walsh-book2,walsh-book1997}
for basic properties and references on Walsh-Hadamard Transforms and \cite{Vetterli2015} for newly-found interesting
properties.

Assume that ${\bf x}$ is corrupted by additive noise ${\bf w}=(w_0,w_1,\ldots,w_{2^{n}-1})$, where 
$w_j$'s are i.i.d. random variables with zero mean and 
variance $\sigma^2$. Note that we \emph{do not} assume that the time-domain components of noise follow the Gaussian distribution.
However, when the signal dimension $2^n$ becomes large,
using orthogonality of columns of Walsh-Hadamard matrix,
we can deduce that the Walsh coefficients of ${\bf w}$ are approximately i.i.d. Gaussian with zero mean and variance
$\sigma'^2 = 2^n \sigma^2$.
Denote the noise-corrupted signal by ${\bf u} = {\bf x} + {\bf w}$ and denote the Walsh-Hadamard transform of ${\bf u}$ by ${\bf y}$.
Define the signal-to-noise ratio (SNR) by
\begin{equation}\label{def_SNR}
\text{SNR} = \frac{\| \widehat{{\bf x}} \|^2}{2^n \sigma'^2}\, ,
\end{equation}
where
$\| \widehat{{\bf x}} \|^2 = \sum_{i=0}^{2^n-1} (\widehat{ x}_i)^2$.
Note that Eq. (\ref{def_SNR}) is consistent with the quantity \cite[Eq.(5)]{arxiv2015wht}. 
The main difference is that 
unlike \cite{isit2015wht,arxiv2015wht} in signal processing,
our definition is not based on the assumption of Gaussian distributions of the time-domain components of the noise.

In signal processing, noisy (sparse) WHT can be very efficiently solved recently (see \cite{isit2015wht,arxiv2015wht}) with
 $\text{SNR} > 0 \text{ dB}$.
Further,
it has been identified that 
\begin{equation}
\text{SNR} > 10\log_{10} \Bigl(\frac{8\log 2}{ 2^n}\Bigr) \text{ dB}
\end{equation}
 has greatest significance in cryptography (and coding theory) 
(see \cite{eprint2016serge,vaudenay_textbook2006,my_arxiv2015}).

\section{Architecture-unaware Parallel WHT}
\label{sec:2}

In this section,
we restrict ourselves to in-place WHT, i.e., the inputs and the outputs are both stored at the same place.
The baseline implementation of WHT is shown in Fig. \ref{algo1_in_memory_wht},
where $LSB_k(\cdot)$ denotes the least significant $k$ bits of the input for $k>0$ and
$LSB_0(\cdot)$ is defined to be $0$.
Our experiences show that this serial version has fairly well performance as long as the required storage does not exceed the available RAM amount.

\begin{figure}[h]
\begin{center}
\begin{algorithmic}[1]
\REQUIRE the time-domain signal $buf[\cdot]$ of dimension $2^n$
\ENSURE the transform-domain signal $buf[\cdot]$
\FOR{$k=0, \ldots, n-1$}
  \STATE $pt\leftarrow 0, j\leftarrow 2^k$ 
  \FOR{$i=0, \ldots, 2^{n-1}-1$}       
     \STATE $temp_1 \leftarrow buf[pt]+buf[pt+j]$
     \STATE $temp_2 \leftarrow buf[pt]-buf[pt+j]$
     \STATE $buf[pt]\leftarrow temp_1$ 
     \STATE $buf[pt+j]\leftarrow temp_2$
     \STATE $pt\leftarrow pt+1$
     \IF{$LSB_k(pt)=0$} 
        \STATE $pt \leftarrow pt+j$
     \ENDIF
  \ENDFOR
\ENDFOR
\end{algorithmic}
\caption{Baseline implementation of $\textsc{WHT}(buf, 2^n)$.}\label{algo1_in_memory_wht}
\end{center}
\end{figure}

To speed up computations, we would like to use multi-cores to parallelize above serial version of WHT. Notably,
WHT is a representative example of data-intensive computing; this becomes the typical work task nowadays.
In general,
we are guided by the two heuristic principles to make our parallel WHT based on Fig. \ref{algo1_in_memory_wht}.
First, try to use data parallelism strategy to split the data computations into $m$ independent subtasks of equal loads for the $m$-core parallel computing.
Second, try to keep the total number of necessary synchronizations among the $m$ running cores small.
For $m=4$, Fig. \ref{algo2_parallel_wht} illustrates this idea.
Here, we managed to split WHT to the serial run of three synchronizations in all.
Each synchronization assigns equal workloads (which are called $subtask$) to $m$ cores to be run in parallel.
Specifically, the first synchronization let each core run WHT with the reduced dimension $2^{n-2}$;
it thus leaves two more rounds to be done in order to complete the original WHT with dimension $2^n$, which can  
be done by two extra synchronizations.
For convenience,
we define the common $\textsc{workload}(\cdot)$ (see Fig. \ref{algo2_parallel_wht}) so that each core can run with different parameters for the last two synchronizations.
Note that each $subtask$ always executes equal amount of computations yet on different part of data, i.e., $buf[\cdot]$.
In the next section, we present large-scale external WHT when the data cannot be held all in the main memory.

\begin{figure}[h]
\centering
\begin{algorithmic}
\REQUIRE the time-domain signal $buf[\cdot]$ of dimension $2^n$
\ENSURE the transform-domain signal $buf[\cdot]$
\STATE $\textsc{workload}(buf_0,pt_0,j_0)$
\STATE \{
    \STATE $buf\leftarrow buf_0, pt\leftarrow pt_0, j\leftarrow j_0$
    \FOR{$i=0, \ldots, 2^{n-3}-1$}
      \STATE $temp_1 \leftarrow buf[pt]+buf[pt+j]$
      \STATE $temp_2 \leftarrow buf[pt]-buf[pt+j]$
      \STATE $buf[pt]\leftarrow temp_1$ 
      \STATE $buf[pt+j]\leftarrow temp_2$
      \STATE $pt\leftarrow pt+1$
    \ENDFOR
    \STATE \}
\REPEAT
\STATE $\text{subtask}_i$: run $\textsc{WHT}(buf+i\cdot 2^{n-2},2^{n-2})$, $i\in [0,3]$ 
\UNTIL {$4$ subtasks are all completed}
\REPEAT
\STATE $\text{subtask}_0$: run $\textsc{workload}(buf,0,2^{n-2})$
\STATE $\text{subtask}_1$: run $\textsc{workload}(buf,2^{n-3},2^{n-2})$   
\STATE $\text{subtask}_2$: run $\textsc{workload}(buf,2^{n-1},2^{n-2})$
\STATE $\text{subtask}_3$: run $\textsc{workload}(buf,2^{n-1}+2^{n-3},2^{n-2})$    
\UNTIL {$4$ subtasks are all completed}
\REPEAT
\STATE $\text{subtask}_i$: run $\textsc{workload}(buf,i\cdot 2^{n-3},2^{n-1})$, $i\in [0,3]$  
\UNTIL {$4$ subtasks are all completed}
\end{algorithmic}
\caption{Example of $\textsc{ParallelWHT}(buf,2^n,m)$ with $m=4$.}\label{algo2_parallel_wht}
\end{figure}

\section{Practical Large-scale External WHT}
\label{sec:3}

\subsection{Performance Modeling}\label{subsec:performance_modeling}

Denote the dimension of WHT by $2^n$.
Let $T$ denote the runtime.
We write it by 
\begin{equation}\label{E_sys_runtime}
T = T_{\text{CPU}} + T_{\text{I/O}}.
\end{equation}
It is the sum of in-memory processing time, denoted by $T_{\text{CPU}}$, and the external I/O processing time,
 denoted by $T_{\text{I/O}}$. 
For reference implementation\footnote{The operating system is Ubuntu~14.04.1,
and the compiler version is gcc~4.8.4 with the optimization flag `-O3'.} on the modern PC equipped with 2.2GHz CPU,
 we will use $T_{\text{CPU}} = 2.5$ seconds for $n=26$.  
Note that $T_{\text{CPU}}$ should always scale regularly with the problem size (herein $2^n$), 
regardless of the implementation details.
For example, with $n=32$, we expect $T_{\text{CPU}} \cong 2^{32-26} \times 2.5 \cong 160$ seconds
for the same implementation choice.

Next, we discuss the factor $T_{\text{I/O}}$ in (\ref{E_sys_runtime}).
Let $T_{\text{cp}}$ denote the time to copy the whole dataset to another
 over the same disk. Given the fixed size of the dataset, $T_{\text{cp}}$ is a stable system-dependent parameter,
and it does not depend on WHT implementation details. We use the disk speed testing tool \verb$dd$.
To test disk writing speed (i.e., not cache writing),
we use:
~\\
\lstinline$dd if=/dev/zero of=bigfile bs=100k count=10k conv=fdatasync$ 
~\\
To test disk reading speed, we use:
\begin{center}
\lstinline$dd if=anotherbigfile of=/dev/null bs=100k count=10k$ 
\end{center}
The local disk is found to have a real reading and writing speed of 
140MB/s, 83MB/s 
 respectively.
This means reading (resp. writing) the 32GB file takes 229 seconds
(resp. 386 seconds).
And our tests of using the system command to copy a 32GB file 
(which corresponds to the size of the dataset for $n=32$),
result in $T_{\text{cp}} \cong 506$ seconds,
which is a lot faster than the sum of $229 + 386 = 615$ seconds (see Sect. \ref{subsec:current_results} for more discussions).

Given $n$ and in-memory processing capability for problem size up to $2^B$,
suppose that external WHT needs $q$-pass of accessing (i.e., reading and writing) the dataset on the disk. 
Then, we have
\[
T_{\text{I/O}} = q \times T_{\text{cp}}.
\]
We can utilize the just-in-memory processing power of $2^B$ to get the optimal $q$ with 
\begin{equation}
q=n-B+1. 
\end{equation}
We give an external WHT in
Fig. \ref{algo_external_wht}, which is directly adapted from in-memory WHT in Fig. \ref{algo1_in_memory_wht}.
It works as follows. After running an initial WHT of the reduced dimension $2^{B}$ 
which can put the whole sub-dataset in memory,
we do not read/write blocks of data at one time.
Instead, we read two entries from the dataset of non-contiguous locations, which are separated by 
distance $j$.
After the butterfly-like calculations,
we then write the two updated entry-values to the previous read-locations respectively.
The next read-pointer is updated.
The whole procedure is iterated and the distance $j$ is increased regularly.
Obviously, the movement of the disk I/O pointer (i.e., to read/write the dataset) 
does not take the optimal strategy with respect to the total jump distance.

For the rotation-based disks, we consider that it is an important factor to the disk I/O performance in general.
We propose two methods to find the optimal strategy for the external WHT using the rotation-based disk(s).
First, perform two separated blocks of reading/writing from disjoint dataset locations.
Vary the block size to find out the optimal size. 
Note that the block size is independent of the parameter $B$.
Secondly, re-arrange different orders of disk I/O operations, calculations for less number and shorter jump distance of
non-contiguous reading/writing.

\begin{figure}[h]
\centering
\begin{algorithmic}[1]
\REQUIRE the time-domain signal $buf[\cdot]$ of dimension $2^n$, the parameter $B$
\ENSURE the transform-domain signal $buf[\cdot]$
\FOR{$u=0, \ldots, 2^{n-B}-1$}
   \STATE read from dataset from index $u\cdot 2^B$ with block size $2^B$ to $buf[\cdot]$
   \STATE run $\textsc{WHT}(buf,2^B)$
   \STATE write back $buf[\cdot]$ into dataset from index $u\cdot 2^B$ with block size $2^B$
\ENDFOR
\FOR{$k=B, \ldots, n-1$}
   \STATE $pt\leftarrow 0, j\leftarrow 2^k$ 
   \FOR{$i=0, \ldots, 2^{n-1}-1$}
       \STATE read from dataset at index $pt$ to $buf_1$
       \STATE read from dataset at index $pt+j$ to $buf_2$
       \STATE $temp_1 \leftarrow buf_1 + buf_2$
       \STATE $temp_2 \leftarrow buf_1 - buf_2$
       \STATE $buf_1\leftarrow temp_1$ 
       \STATE $buf_2\leftarrow temp_2$
       \STATE write back $buf_1$ into dataset at index $pt$
       \STATE write back $buf_2$ into dataset at index $pt+j$       
       \STATE $pt\leftarrow pt+1$
       \IF{$LSB_k(pt) = 0$} 
           \STATE $pt \leftarrow pt+j$
       \ENDIF 
   \ENDFOR 
\ENDFOR
\end{algorithmic}
\caption{An external WHT directly adapted from Fig. \ref{algo1_in_memory_wht}.}\label{algo_external_wht}
\end{figure}

\subsection{Current Results}\label{subsec:current_results}

\begin{figure}[h]
\centering
\includegraphics[scale=.5]{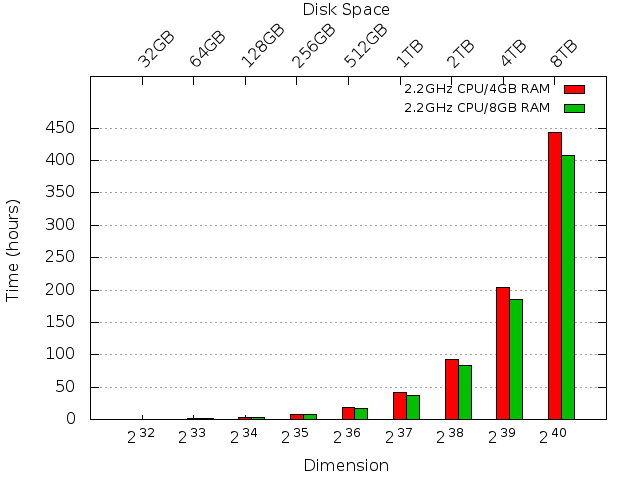}
\caption{Reference runtime for external WHT on the PC with 2.2GHz CPU, 16GB RAM.}\label{table:1c}
\end{figure}

\begin{figure}[h]
\centering
\subfloat[][Running on PC with 2.2GHz CPU, 4GB RAM]{\includegraphics[scale=.5]{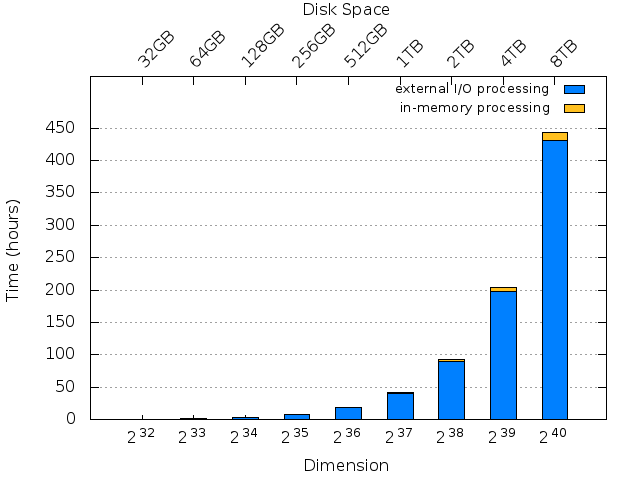}\label{table:1c_detailed_view:a}}
\medskip
\subfloat[][Running on PC with 2.2GHz CPU, 8GB RAM]{\includegraphics[scale=.5]{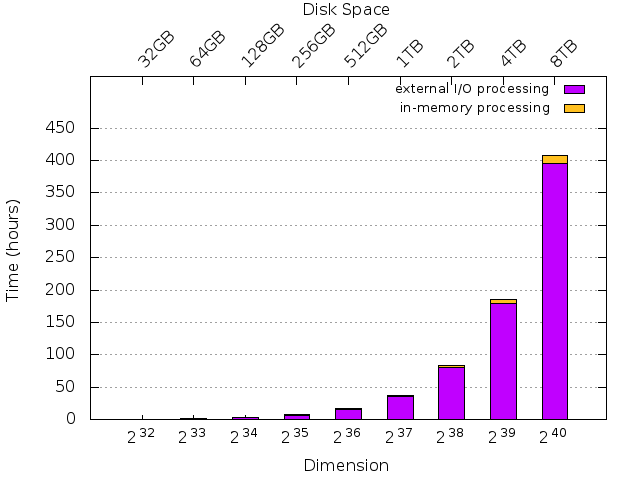}\label{table:1c_detailed_view:b}}
\caption{Detailed runtime diagnostics for external WHT (in Fig. \ref{table:1c}).}
\label{table:1c_detailed_view}
\end{figure}

For $n=32$, we use $B=30$ on our PC, and we get $q=3$, $T_{\text{I/O}} \cong 3\times 506 = 1518$ seconds.
It makes $T \cong 160 + 1518 = 1678 $ seconds (about 28 mins).
We see that $T$ is dominated by the factor $T_{\text{I/O}}$ rather than $T_{\text{CPU}}$.
Fig. \ref{table:1c} gives the reference runtime on the PC with 2.2GHz CPU and 16GB RAM using 4GB, 8GB RAM respectively 
(i.e., corresponding to $B = 29,30$).
Notably, performing \emph{truly tera-scale} WHT (i.e., with the dimension $2^{40}$) 
can be done within 3 weeks %
using 8TB disk space\footnote{In this case, multiple disks might be required.}.
It is worth pointing out that the required RAM amount does not make the performance improvement in proportion.
Fig. \ref{table:1c_detailed_view} shows the detailed runtime diagnostics.
The in-memory processing requires the same amounts of time;
 the external I/O processing time is the dominant factor.

\begin{figure}[!t]
\begin{center}
\includegraphics[scale=.5]{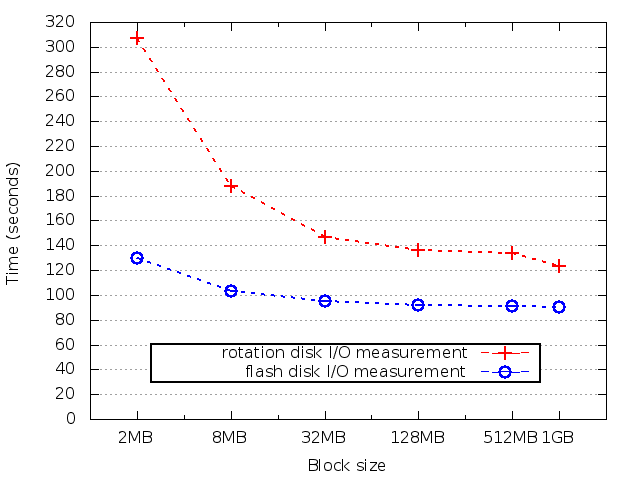}
\caption{Low-level I/O measurements: the total time to copy a different 8GB file decreases
by increasing the block size of internal transfer.}\label{fig_block_test}
\end{center}
\end{figure}

In our experiments with $B=29,n=32$,
the runtime 2468 seconds is obtained, in contrast to the reference time 0.61 hours in Fig. \ref{table:1c},
which is obtained by $(160 + 4\times 506) = 2184$ seconds.
By detailed analysis, we found that for one pass, the real $T_{\text{I/O}}$ is always around $577$ seconds,
while the theoretical $T_{\text{I/O}}$ is $506$ seconds.
This difference accounts solely for the total runtime difference of around
$4\times (577-506) = 284$ seconds.
Recall as we have mentioned in Sect. \ref{subsec:performance_modeling} that the system copy command (for a 32GB file)
takes 506 seconds,
which is much faster than the sum of reading the file and then writing a file of equal size.
To solve this problem, we decide to write low-level I/O tests in order to know more accurate time of copying a file.
For the block size of 2MB, 8MB, 32MB, 128MB, 512MB respectively,
we copy a different file of 8GB separately and measure the total time.
In our tests,
we use the direct I/O writing to eliminate cache writing effects.
The results are plotted in Fig. \ref{fig_block_test} (in red).
Note that when we tried with a larger block size of 2GB as the system allows physically, 
the file reading operation fails each time we run.
So, we attempted with a smaller block size of 1GB.
Our experiences show that a block size ranging from 128MB to 1GB is amenable for the rotation-based disk I/O.

\subsection{Faster Portable External WHT}

It is clear that the rotation-based disks suffer severely data-intensive computing. 
Our results with the 7200RPM Seagate disk shows that its performance is far from satisfactory,
considering specifically the fact that the SATA6.0 interface is supposed to offer a much higher throughput, that is, 6Gb/s corresponding to 750MB/s.

We decided to experiment with the USB3.0 flash disk which offers more than 200MB/s speed for both read and write by the specification.
For the flash disk, we did the same low-level I/O tests to know more accurate time of copying a file as done before.
The results are shown in Fig.~\ref{fig_block_test} (in blue).
We notice two remarkable facts.
First, both the rotation disk and the flash disk exhibit very similar performance,
i.e., the total file copy time decreases and converges with increasing block sizes.
Interestingly,
the larger block size 2GB always fails with file reading.
And we consider that it is the internal memory I/O fault rather than the external I/O fault that causes the trouble.
This implies 1GB block size the best.
For highly data-intensive computing, it is suggested to choose the block size 128MB to 512MB to guarantee error-free I/O operations.
Secondly,
for each block size we have tested, the flash disk improves the performance over the rotation disk by a fairly stable constant factor
of around $1/3$.
We have extensive experiences with rotation disks over various workstations.
The best I/O performance 
these disks can offer
under our focused data-intensive computing model
 seems to be no better than our results in Fig.~\ref{fig_block_test}.
On the other hand,
we find that various choices of business-class flash disks are available,
which offer different rates between reading and writing.
In this work, we chose a flash disk with a balanced rate, 
which claims an official reading and writing rate of 260MB/s, 240MB/s respectively.

In general, the flash disk has two different properties compared to the rotation disk with respect to the performance:
the former has much quicker access time than the latter;
the former is insensitive to the jump distance between two non-contiguous access regions,
which is not the case for the latter. 
It is therefore expected that we can further improve the performance of external WHT.

\begin{figure}[!t]
\begin{center}
\includegraphics[scale=.5]{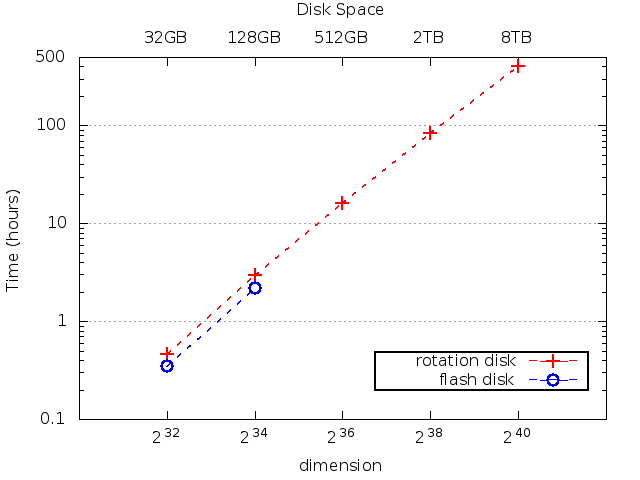}
\caption{Optimal External WHT.}\label{fig_optimal_ext_wht}
\end{center}
\end{figure}

Finally, we give the optimal performance results of the external WHT in Fig.~\ref{fig_optimal_ext_wht}. 
Current USB flash technology can support high transfer rate of large capacity ranging from 128GB, 256GB, 512GB.
This limits the dimension of portable external WHT to smaller than $2^{36}$.
Nevertheless, 
we see that high-performance flash disks always beat rotation disks whenever the disk space allows.
Particularly, giga-scale/tera-scale WHT
 has notable cryptographic significance (cf. \cite{my_new_journal_article2015}).
With the dimension $2^{32}$,
using 4GB and 8GB RAM,
one can run WHT with
time 2160 seconds,
1660 seconds respectively over the rotation disk;
on our flash disk,
this can be done with time 1628 seconds,
1261 seconds respectively.
As another example,
with the dimension $2^{35}$,
using 8GB RAM and 256GB disk space, WHT can be done with time
7 hours, 5.3 hours
over the rotation disk, the flash disk respectively.
Currently, we are experimenting with WHT over practical large distributions.
Our first target distribution counts the number of prime factors for each natural number.
The results will be available in the final version \cite{my_ftc2016_full_version} of the paper.

\section{More Results on Practical Tera-scale WHT}
\label{sec:4}

According to our results,
we can immediately build practical systems to efficiently compute WHT of dimension larger than $2^{40}$.

\subsection{WHT with On-the-fly Signal Source}
\label{subsec:4_1}
Suppose that each PC can have independent parallel on-line access to the signal source.
This is the case when we have a digital signal source of some cryptographic function for example.
We note that the dominant part of computations is the external I/O. 
For $n=2^{40}$, the optimal external I/O and the in-memory processing time takes time around
 $(5500\times 2^{8})$ seconds, $(160\times 2^{8})$ seconds
 on the rotation disk respectively, according to our analysis in Sect. \ref{sec:3}.
So, trying to increase the in-memory processing time and maintain the external I/O loads
will not lose much in performance.

Thus, we propose to use multiple PCs of same settings to compute WHT of dimension $2^{45}$ as follows.
Let each PC receive the same copy of the time-domain signal on-the-fly (and we do not consider the time cost for it).
We want the Walsh coefficients to be calculated and stored in a distributed way.
Each PC calculates Walsh coefficients over a linear subspace of the original source:
 it pre-processes the dataset of its own copy by choosing a random linear space reduction $GF(2)^{45} \to GF(2)^{40}$ 
and then stores an initial sub-dataset of size $2^{40}$ to the external disk(s).
Next, each PC does WHT with the initial sub-dataset and reduced dimension $n=2^{40}$.
Suppose that each PC uses 8GB RAM and 8TB hard disk space.
We estimate the total time cost of each PC by 
$(160\times 2^{8+5} + 5500\times 2^{8})$ seconds, which is approximately $32$ days.
Hence, each PC computes a portion of $1/32$ of all Walsh coefficients.
Using $64$ PCs, approximately $1 - (31/32)^{64} \approx 87\%$ of all Walsh coefficients are obtained within $32$ days.
This beats our previous external WHT record based on rotation disks 
(see Fig. \ref{fig_optimal_ext_wht} in Sect. \ref{sec:3}),
which can handle WHT with dimension no larger than $2^{40}$.
We note that this technique can be further optimized by different parameter choices in order to obtain all Walsh coefficients with probability close to 1.

\subsection{Advanced Techniques}

In Sect. \ref{sec:2}, we have presented a parallel WHT for
a system without cache. 
It is intended to speed up in-memory WHT by multi-cores, 
and thus it is not helpful when the dimension becomes larger.
Now, we discuss advanced techniques to extend parallel WHT to larger dimensions.
We are interested in the general case that the signal source is not on-the-fly.

\begin{figure}[h]
\centering
\subfloat[][]{\includegraphics[scale=.45]{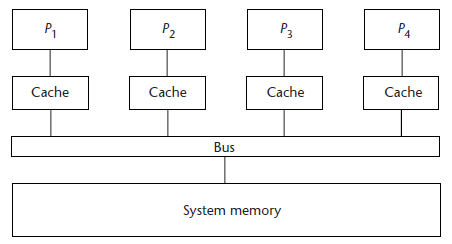}\label{fig_HPC_architectures:a}}
\qquad
\subfloat[][]{\includegraphics[scale=.45]{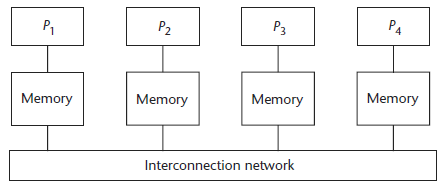}\label{fig_HPC_architectures:b}}
\caption{Two HPC architectures: the shared-memory systems (Left) and the distributed-memory systems (Right).}
\label{fig_HPC_architectures}
\end{figure}

In High-Performance Computing (HPC), there exist two
different systems to solve extremely large scale problems (cf. \cite{supercomputing_book2005}). They are the
shared-memory systems and the distributed-memory systems (see Fig. \ref{fig_HPC_architectures}). The former
has a single shared address space (i.e., main memory) that can be accessed
by any processor. For the latter, the system memory is packaged 
with individual nodes of one or more processors, and communication is
required to access data from one processor's memory by another processor
via the interconnect interface.

We use a model of the shared-memory system architecture\footnote{It  
 is often associated with SMP, which stands for Shared-Memory Processing/Programming.} as follows.
Like its HPC counterpart, it has a single shared address space (i.e., main memory) that can be accessed by any core.
Suppose that the system consists of two-level caches, namely, L1 cache and L2 cache.
Each core has both an L1 (data) cache and a bigger L2 cache of its own.
We choose the SMP architecture with the non-shared last-level cache as our best multi-core platforms.
Note that though elaborate fine-tuned techniques (such as machine-level vectorization, latency hiding) can be used to improve the parallel performance in Sect. \ref{sec:2}, to parallelize WHT on multiple computing nodes,
data transfer rate between multiple nodes becomes the main bottleneck.
Based on our previous analysis, 
we need the node-to-node data transfer rate many times higher than the data transfer rate that the local storage can afford;
otherwise, we can simply make each node run the reduced computing task independently as discussed in Sect. \ref{subsec:4_1}.
Obviously, the typical network setting of 1GbE (which stands for Gigabit Ethernet) is insufficient.
While the cost of switching to 10GbE network is still broadly prohibitive,
we propose a custom solution to use multiple GbE network cards for each node.
This way, we can make full use of several nodes to do parallel computing.
We will report the MPI implementation details in the final version \cite{my_ftc2016_full_version} of the paper.

\section{Summary}\label{sec:end}

WHT is popular in a variety of scientific and engineering applications due to 
its power and simplicity.
Most recently, interdisciplinary research efforts and interests emerge that center around the topic of (noisy) sparse WHT.
In this work,
we study efficient implementations of very large dimensional general WHT.
Our results show \emph{for the first time} that WHT of dimension $2^{35}$ can be done within a quarter of a day;
WHT of dimension $2^{40}$ can be done within around 400 hours.
Undoubtedly,
WHT plays an important role in noisy sparse WHT,
as the former can be seen as the first solution to the latter especially when the dimension is large and the noise is strong.
Our work is believed to shed light
on noisy sparse WHT, which remains a big open challenge in academia.
Further, the main idea behind also helps to study parallel data-intensive computing, 
which has a broad range of applications.

\section*{Acknowledgment}

This paper was supported by the Norwegian Research Council under project number 247742/O70.

\end{document}